\documentclass[aps,pre,twocolumn,groupedaddress,showpacs]{revtex4}

\usepackage{graphicx}
\usepackage{color}

\begin{document}
\newcommand{\dc}{$^\circ$C}


\title{Rotation of an ice disc while melting on a solid plate}


\author{S. Dorbolo, N. Vandewalle, and B. Darbois-Texier}
\affiliation{
CESAM-GRASP, D\'epartement de Physique B5, Universit\'e de Li\`ege, B-4000 Li\`ege, Belgium.\\
}


\date{\today}

\begin{abstract}
Ice discs were released at the surface of a thermalized aluminium plate. The fusion of the ice creates a lubrication film between the ice disc and the plate. The disc becomes very mobile. The situation is similar to the Leidenfrost effect reported for liquid droplet evaporating at the surface of a plate which temperature is over the boiling temperature of the liquid. For the ice discs, we observe that, while the ice discs were melting, they were rotating.   The ice disc starts rotating. The rotation speed increases with the temperature of the plate and with the load put on the ice disc. A model is proposed to explain the spontaneous rotation of the ice disc. We claim that the rotation is due to the viscous entrainment of the ice disc by the liquid that flows around the ice disc. \end{abstract}

\pacs{64.70.dj,47.85.mf}

\maketitle

\section{Introduction}
The levitation of a liquid droplet can be obtained by dropping the droplet on a plate which temperature is above the boiling temperature of the liquid \cite{quere}. The evaporisation is that efficient that a thin lubrication film is established between the droplet and the plate. That is the so-called Leidenfrost effect. In Ref. \cite{clanet}, a block of dry ice put on a hot plate levitates on a cushion of carbon dioxide gas. The solid sublimates into gas which produces the intervening lubrication film.  The sublimation Leidenfrost was recently used to put in rotation a disc of dry ice on a textured plate making a genuine turbine \cite{machale}. Generally speaking, the sustentation is possible because of a change of phase; the energy is provided by the plate or by the droplet.

In this paper, we are interested in a third transition which allows levitation, i.e. the fusion. Ice discs were put on a thermalised aluminium plate. While melting, a lubrication film forms between the plate and the disc. The disc is very mobile and even starts rotating. We related this phenomenon to the rotation of ice disc at the surface of a thermalised bath and the model we propose here is similar to what was developed in Ref.\cite{us}.

We propose a series of experiments that investigate the behavior of the melting ice block. First, the experimental set-up is described in Sect. II. Afterwards, the experimental results are presented in Sect. III. We start by the determination of the melting rate as a function of the plate temperature. Afterwards, the rotation speed of the disc in function of the plate temperature and in function of the mass loaded on the ice disc is determined. The liquid flows are then revealed by coloring the water surrounding the ice disc. The model is described in two parts in Sect. IV: (i) discussion about the liquid film locating between the ice disc and the plate and (ii) the influence of the liquid flows on the rotation speed of the ice disc. Finally, conclusions are drawn in Sect. V.
\section{Experimental set up}

\begin{figure}[!h]
\centering
(a) \includegraphics[width=8cm]{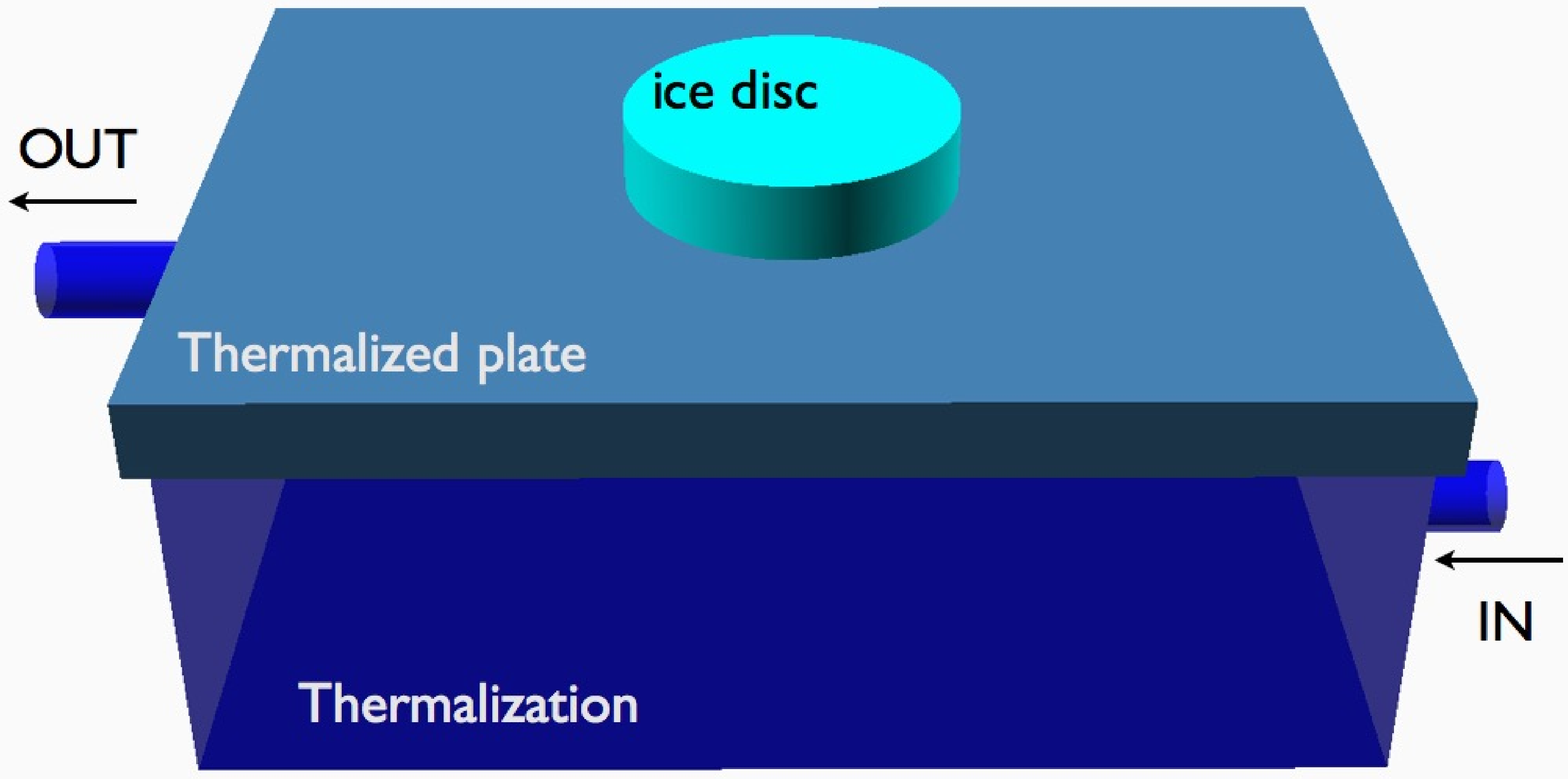}

(b) \includegraphics[width=8cm]{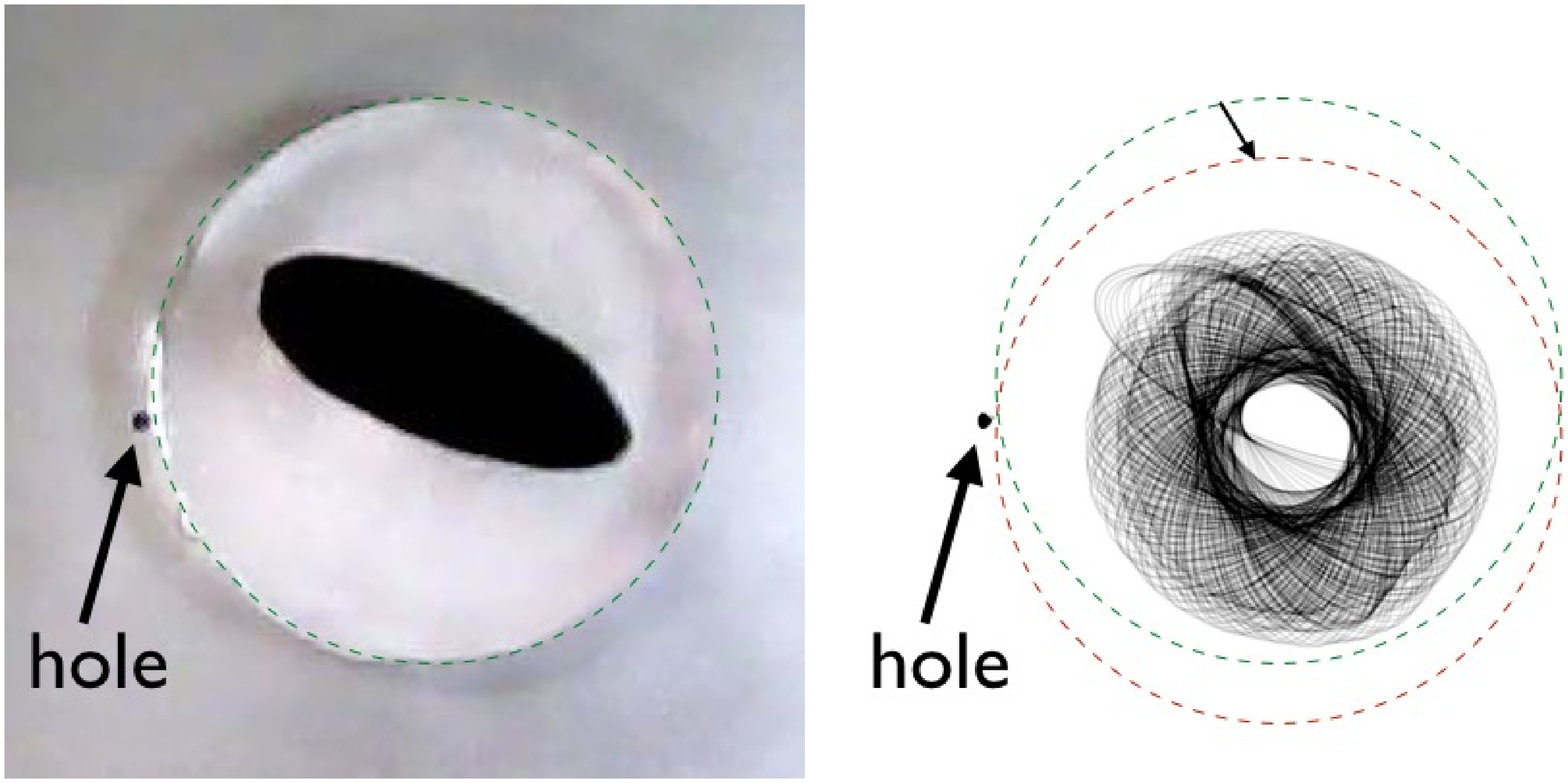}
\caption{(a) Sketch of the experimental set-up. An ice disc is placed on an aluminum plate in light blue. Thermalized water circulates in the thermalization box. (b) On the left, a picture of the ice disc is shown from the top. The evacuation hole is on the left of the disc and the black ellipse is visible at the center of the picture. On the right, the superposition of the positions of the ellipse is shown (1 second separates two successive pictures). The broken circles represent the position of the ice disc. The arrow indicates the small motion of translation that occurs, in the present case, at the beginning of the melting. }
\end{figure}

The ice discs were melted on an aluminium plate ( 375 $\times$ 260 $\times$ 8 mm$^3$). A 2 mm hole was drilled in the center of the plate and allows melted ice to flow out of the bottom of the disc. This hole avoids that the ice disc eventually floats on its own liquid. The aluminium plate constituted the top cover of a thermal isolated box ( 375 $\times$ 260 $\times$ 130 mm$^3$) through which circulated thermalised water from a Julabo bath thermalizer (see Fig.1a). As the box contained about 10L of thermalised water, the plate temperature remained constant during the melting of the disc. This system allowed to set the temperature $T$ of the plate  between 4 and 35\dc. 
\begin{figure}[!h]
\centering
(a)\includegraphics[width=8cm]{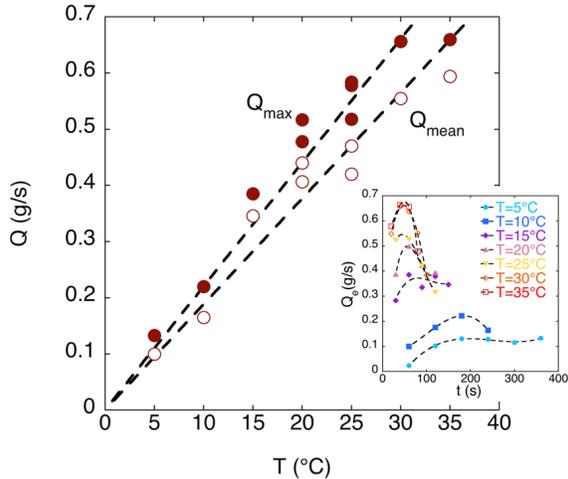}

(b)\includegraphics[width=8cm]{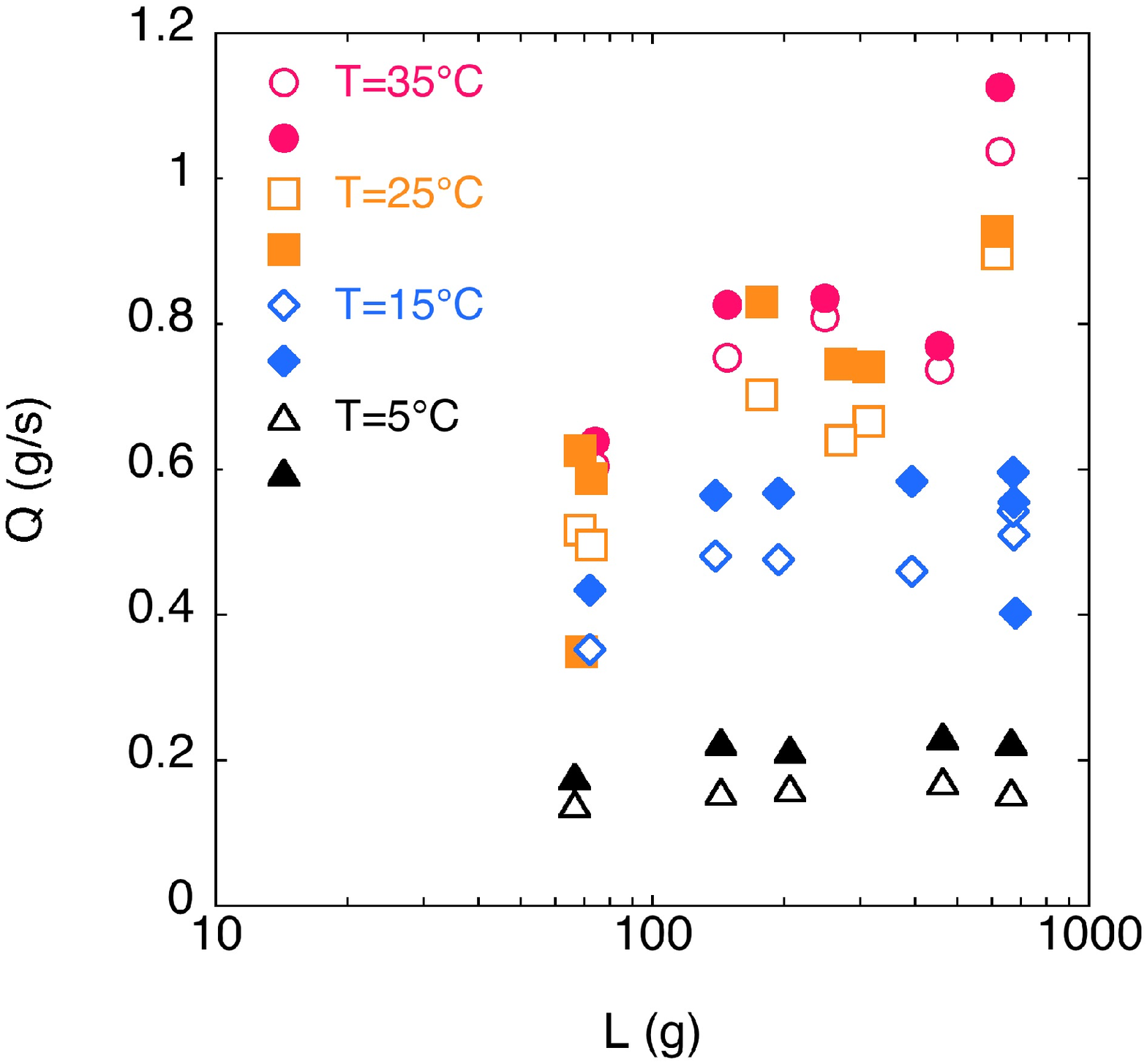}
\caption{(a) Maximum (bullets) and mean (open circles) melting rate of the ice disc as a function of the temperature of the plate.  (inset) Evolution of the experimental melting rate in function of time for different temperature from 5\dc  to 35\dc$\ $   by step of 5\dc  (see Legend). (b) Maximum (plain symbols) and mean (open symbols) melting rate of the ice disc as a function of the load $L$ at four temperatures 5\dc(triangles), 15\dc(diamonds), 25\dc(squares) and 35\dc(bullets).}

\end{figure}

The plate was coated by petroleum jelly. This allows to obtain reproducible wetting conditions. The rest angle of a droplet has been found to be 80$^\circ \pm 6$, the receding and advancing angle 20$^\circ \pm$5 and 119$^\circ \pm$ 6 respectively. The large difference between the receding and advancing angles allows to consider that the contact line is well pinned at the surface.

The ice discs were produced out of circular Petri dish (42.5 mm of radius $R$ and 14 mm of height $h$). The Petri dishes were embedded in polystyrene. Only the bottom part of the petri dish was not covered.  The polystyrene frame containing petri dish filled with water were placed in a freezer (-32\dc).  The water then started freezing by the bottom of the Petri dish, bubbles and dust were pushed towards the surface of the water.  We obtained bubble free ice discs.
The ice discs were placed on the plate. A dark ellipse (50 mm long and 20 mm large) was put in the center of the ice disc. By image analysis, the position of the discs was determined thanks to the detection of the dark ellipse after thresholding the image. The images were captured with a webcam and analysed in real-time using a Python program and OpenCV. An example of the analyzed pictures is shown in Fig. 1b (on the right, the image and on the left the superposition of successive analyzed images).

\section{Results}
The results sections is separated in two parts. First we measured the melting rate of the ice discs as a function of the temperature of the plate. Afterwards, we measured the rotation speed as a function of the temperature. Finally, we loaded the ice in order to check the influence of the pressure on the rotating speed.
\subsection{Melting rate}
The melting rate was estimated by weighting the ice disc after time steps. The operation was performed for plate temperature $T$ ranging between 5 and 35\dc$\ $by steps of 5\dc. The time between two successive weight measurements was $t_s=$60 s when $T=$5, 10\dc, $t_s=$30 s when $T=15$, 20\dc, and $t_s=20$ s when $T=25$, 30 and 35\dc. The experimental flow rate is then given by $Q_e=(m(t)-m(t-t_s))/t_s$. The measurements are presented in the inset of Fig.2a.  As noticed, the hotter is the plate, the faster the ice disc is melting. Moreover, one observes a maximum in the experimental flow rate which is more marked when the temperature of the plate is high. This effect is due to the time to establish the liquid film between the disc and the plate. The maximum flow rate $Q_{max}$ (bullets) and the mean flow rate $Q_{mean}$ (open circles) have been reported as a function of the plate temperature in Fig.2a. One finds that the phenomenological trend is linear. The flow rate can be modelled, in first approximation, as following the empirical law: $Q=A T$ where $A=$20.10$^{-6}$ kg s$^{-1}$ K$^{-1}$. 

The mass dependence of the melting rate has been measured for four temperatures (5\dc, 15\dc, 25\dc$\ $and 35\dc) and this for loads $L$ between 60 and 600 g. The load $L$ is given by the mass of the ice disc plus an additive mass. The maximum and the mean flow rates are reported in Fig.2b. One finds that the flow rate dependence with the load first increases with small load (until about 150 g). For larger loads, the melting rate is rather not influenced by the load, except for the highest temperature for which an increase of about 50\% of the melting rate is observed.

\subsection{Rotation speed}
The rotation speed of ice discs ($L \approx 70$ g) was measured as a function of the plate temperature. The data is reported in Fig. 3a. We consider that when the temperature of the plate is 0\dc, the rotation speed is also 0.  When the plate temperature $T$ is increased, the rotation is also observed to increase. Such a behavior is clearly related to the increase of the melting rate with the plate temperature.

In addition, we loaded the ice disc with a Petri dish containing lead spheres. This allows to increase the pressure on the liquid film. The rotation speed of the ice disc is presented in function of the load $L$ (ice disc mass plus charge) for a plate temperature equal to 20\dc. The general trend is the increase of the rotation speed with the load $L$.

\subsection{Flow vizualisation}
During the ice melting at $T=$20\dc, some dye was added around the melting ice disc. This experiment allowed visualizing the flow around the ice disc. The flow around the disc escaped from the plate by the small evacuation hole.  One found that part of the flow rotated in the same direction as the rotation of the disc while the rest of the flow rotated in the opposite direction.  Consequently there exists a point of divergence for the flow on the circumference of the disc.  Movies are shown in the supplemental materials \cite{sm}

The different observations show that the liquid flow around the ice disc is responsible for the rotation. This flow occurs in a thin water rivulet bounded by three lines: (i) the contact line between the disc and the plate, (ii) the contact line between the rivulet and the ice disc, and (iii) the contact line between the rivulet and the plate. The proportion of the liquid rotating clockwise or counterclockwise determines the sense of rotation of the disc.

\begin{figure}[!h]
\centering
(a) \includegraphics[width=8cm]{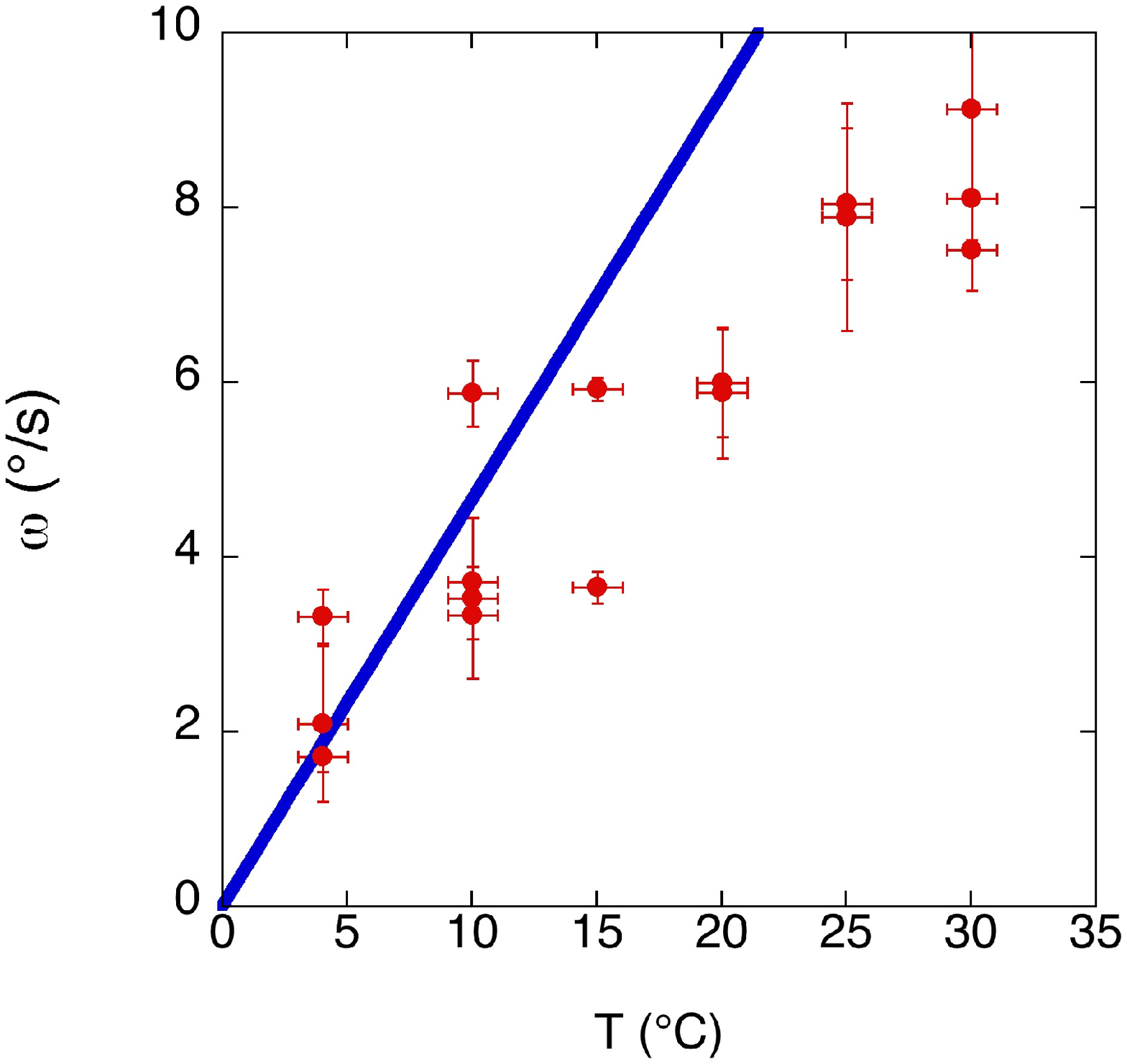}

(b)
\includegraphics[width=8cm]{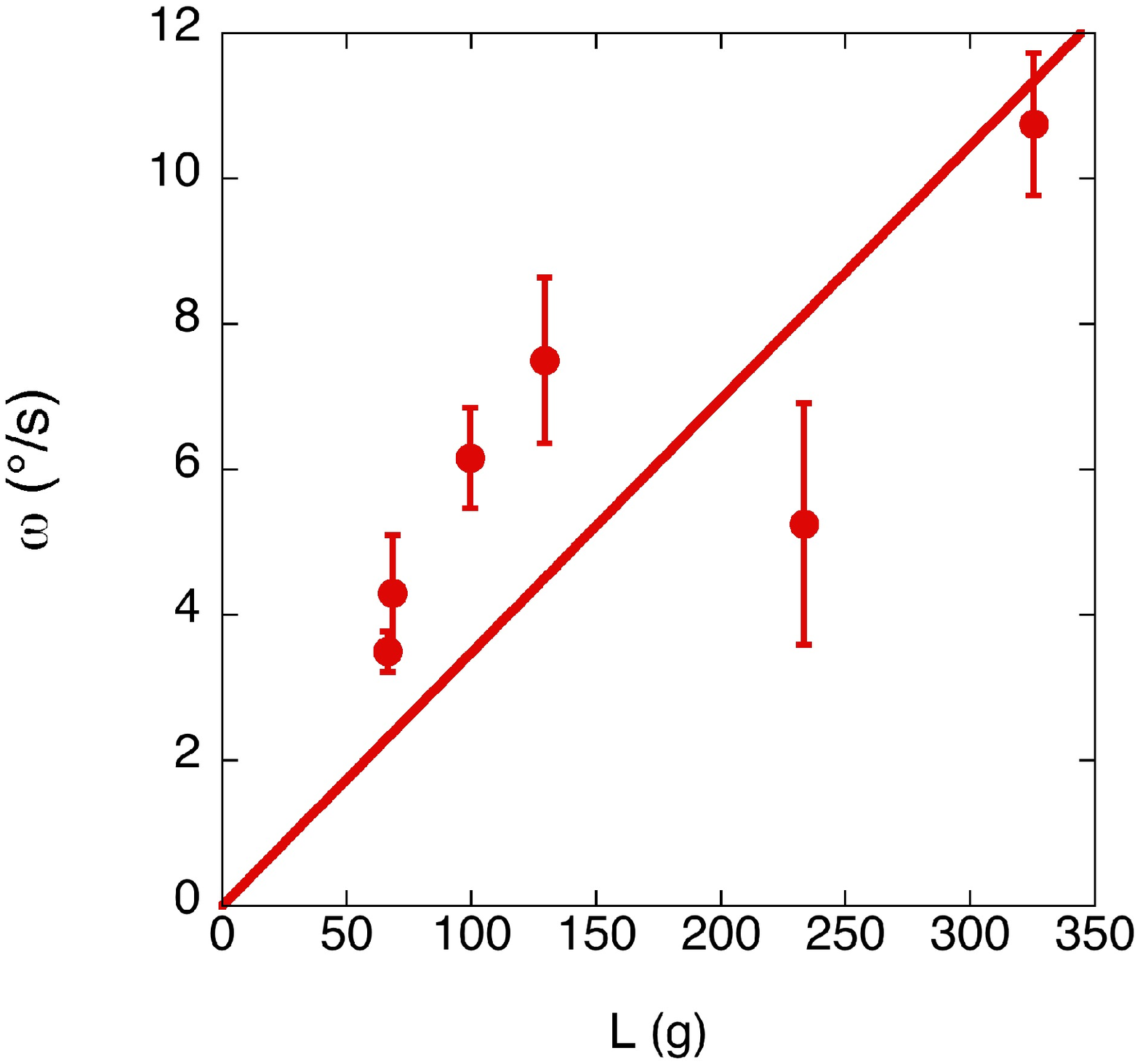}

\caption{(a) Rotation speed $\omega$ (in $^\circ$/s) as a function of the plate temperature. The ice disc masses were about 70 g. The solid line is given by Eq.(10) by taking $a^\star$=0.88. (b) Rotation speed $\omega$ (in $^\circ$/s) as a function of the load (ice disc+charge). The temperature of the plate was $T=$25\dc.}
\end{figure}
\section{Model and interpretations}
\subsection{Liquid film thickness}
The liquid film thickness $e$ between the ice disc and the plate can be evaluated by considering the heat flux (by conduction solely) necessary to melt the ice at the measured melting rate $Q$. In a first step, we took the equation that applies to Leidenfrost droplet, namely
\begin{equation}
Q=\frac{\kappa_w(T=0^oC)}{\mathcal{L}}\frac{\pi R^2}{e}\Delta T
\end{equation}where $\kappa_w$ is the thermal conductivity of the water at 0\dc (=0.55 W m$^{-1}$ K$^{-1}$), $\mathcal{L}$ is the latent heat of fusion of the ice. The melting flow rate $Q$ has been experimentally found to be equal to $A T$ where $A=20.10^{-6}$ kg K$^{-1}$ (Fig. 2a). As the temperature is measured in \dc,  $T$ and $\Delta T$ have the same value and the thickness is found to be independent on the temperature. Computing Eq.(1), one finds $e\approx$500 $\mu$m.
Another way to determine the liquid film thickness is to consider the flow description under the ice block. The mean velocity $\bar v$ of the flow at the edge of the ice disc is given by  
\begin{equation}
\bar v=\frac{Q}{\rho_w 2 \pi R e}\end{equation} where $\rho_w$ is the density of the water at 0\dc. On the other hand, the mean velocity can be also determined by computing the flow under the ice disc in the lubrication approximation. Under these conditions, the flow results from the pressure $p$ of the ice disc on the liquid film. In the system of coordinates where the liquid film spreads in the $x$-$y$ plane and the $z$ direction is perpendicular to the ice disc, the lubrication equation reads
\begin{equation}
\frac{\Delta p}{R}=\eta_w \frac{d^2 \bar v(z)}{dz^2}
\end{equation}where $\eta_w$ is the viscosity of water. The pressure gradient $\Delta p$ is due to the ice disc weight divided by the basis area of the disc, namely $\Delta p=h \rho_i g$ where $h$ is the height of the ice disc and $\rho_i$ the density of the ice, $g$ being the Earth gravity. We solve Eq.(3) under two boundary conditions: (a) $\bar v(z=0)=0$ and $\bar v(z=e)=0$ (Poiseuille) and (b) $\bar v(z=0)=0$ and $\bar v(z=e)$ is maximum (semi-plug flow). One finds
\begin{equation}
\bar v=\beta \frac{\rho_i g h}{\eta R}e^2
\end{equation} where $\beta=$1/12 (Poiseuille case) and 1/3 (semi-plug case). Balancing Eq.(2) and Eq.(4), one finds the expression of the liquid film thickness
\begin{equation}
e=\sqrt[3]{\gamma \frac{\eta_w Q}{\pi \rho_i \rho_w h}}
\end{equation} with $\gamma$=6 or 3/2 in the Poiseuille and semi-plug flow cases respectively. When the temperature of the plate is increased the film thickness increase as $T^{1/3}$. The relation Eq.(5) also indicates that the liquid film thickness increases with time as the height $h$ of the ice block decreases. This explains the melting rate $Q$ (see inset Fig. 2a) decreases with the time as predicted by Eq.(1) when $e$ increases. Finally, the liquid film thickness is found to be equal to 185 $\mu$m and 116 $\mu$m at $T=$20 \dc $\ $ for the Poiseuille and semi-plug flow respectively. This contrasts with the approximation of the mean flow rate used to find the liquid thickness from Eq.(1). Finally, when the thickness of the liquid film is smaller than the capillary length $\kappa=\sqrt{\sigma/\rho_w g}$ (where $\sigma=72$ mN/m is the surface tension of the water), the ice disc starts floating on its own melted liquid. The pressure exerted on the liquid film is reduced to zero.

\subsection{Rotation mechanism}
We consider an ice disc rotating counterclockwise with a constant angular speed $\omega$. The proposed mechanism resides on the entrainment of the disc by the water flows along the perimeter of the disc. This flow is constrained in a rivulet that is bounded by the contact lines along the ice block and along the plate. The viscous entrainment due to the flow counterclockwise $\Phi_+$ is the motor of the rotation. However, as shown in the experiments, a part of the flow turns clockwise $\Phi_-$. The budget between both opposite flows determines the sense of rotation of the disc. In addition, the friction of the disc with the plate has to be taken into account. The balance between the motor (viscous entrainment by $\Phi_+$) and the braking mechanism (viscous entrainment by $\Phi_-$ and friction disc-plate) determines the value of the constant rotation speed of the ice block.

The angles $\alpha$ around the disc are counted positively counterclockwise; $\alpha=0$ corresponds to the position of the evacuation hole  (see Fig.4). The divergence point that separates the counterclockwise and the clockwise flow is located at $\alpha^\star$.  For more convenience, we define the parameter $a^\star=\alpha^\star/2 \pi$. The ice disc melts at a rate $Q=A T$ as found experimentally with $A=20.10^{-6}$ kg s$^{-1}$ K$^{-1}$.  The liquid is evacuated out of the ice disc and flows in a rivulet that surrounds the basis of the ice disc. The section of the rivulet is modelled by a triangle which basis and height are equal to the capillary length $\kappa$.

\begin{figure}[!h]
\centering
\includegraphics[width=8cm]{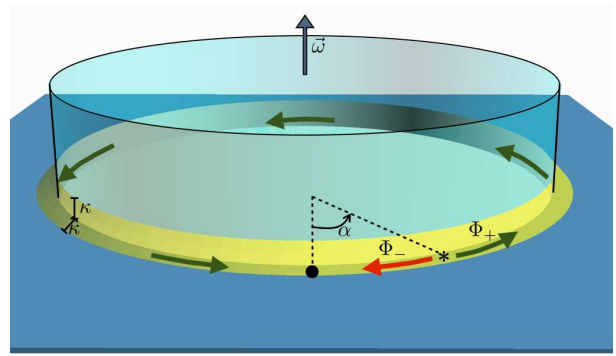}

\caption{(color on line) Sketch of the melting ice disc. The circular rivulet (in yellow) follows the contact line of the ice disc and the plate. The flow counterclockwise and clockwise are represented by green $\Phi_+$ and red $\Phi_-$ arrows respectively. The divergence point is indicated by a star. The hole is shown in front of the ice disc (black bullet).}
\end{figure}
The hypothesis is that the viscous entrainement of the ice disc by the flow in the rivulet is responsible for the rotation of the disc. The viscous entrainement is given by
\begin{equation}
f=\frac{\ell h_e \eta_w}{\delta}v_l
\end{equation} where $\delta$ is the thickness of the boundary layer, $\ell$ is the length of the rivulet, $h_e$ the height of the contact between the rivulet and the disc and $v_l$ the speed of the liquid flow. Applied to the present case $h_e$ is proportional to $\kappa$. 

The counterclockwise flow $\Phi_+$ flows in the rivulet on a length $\ell_+=2 \pi R (1-a^\star)$ while $\Phi_-$ flows in the rivulet on a length $\ell_-=2 \pi R a^\star$.  The melting flow rate $Q$ is distributed among both flows $\Phi_+$ and $\Phi_-$.  Expressed in volume flow, one finds $\Phi_+=a^\star Q /\rho_w$ and $\Phi_-=(1-a^\star) Q /\rho_w$. The mean velocity of the flow can be deduced $v_+=\Phi_+/2S$ and  $v_-=\Phi_-/2S$ where $S=\kappa^2/2$ is the surface of the cross-section of the rivulet. The resulting force due to the viscous entrainement is then given by
\begin{equation}
f_v=a^\star \frac{2\pi R\kappa \eta_w}{\delta}(v_+-\omega R) - (1-a^\star) \frac{2\pi R\kappa \eta_w}{\delta} (v_-+\omega R)
\end{equation} The friction $f_f$ of the ice disc with the plate is also due to the viscous friction, namely
\begin{equation}
f_f=\int_0^R \int_0^{2\pi}r dr d\theta \eta_w\frac{\omega r }{\delta}=2\pi \eta_w \frac{\omega R^3}{3 \delta}
\end{equation} Balancing Eq.(6) and Eq.(8), one finds an expression for the angular speed of the ice disc
\begin{equation}
\omega=\frac{3 \kappa}{R^2/3+R \kappa} \left (a^\star v_+- (1-a^\star) v_- \right )
\end{equation} Replacing the values of $v_+$ and $v_-$ and $Q=A T$, one obtains
\begin{equation}
\omega=\frac{3}{\rho_w \kappa (R^2+3R \kappa)}(2 a^\star-1) A T
\end{equation} The angular speed is found to be proportional to the temperature. The factor $a^\star$ has been estimated experimentally (see Sect. IV); $a^\star$=0.6. On the other hand, there exists a special value $a^\star_0$ for which $v_-=\omega R$. This later relation occurs when $a^\star$=0.88. Taking into account this value, we reported in Fig. 3a the relation Eq.(10). This is the solid line. The agreement with the experiment is satisfactory.

According to this model, the dependence of the rotation speed with the load $L$ can only be found in the melting rate. The reported measurement in Fig. 3b shows an increase of the rotation speed. At 25\dc, the rotation speeds is about 4 deg/s for a load of 70g and is doubled when the load is 300 g. The melting rate measurements are less clear due to the manner it was measured. However, we can see an increase of the melting rate. The melting rate 0.6 g/s for a load of 70 g and nearly 1 g/s when the load is 600 g. The relation between the melting rate and the load is not linear. However, the tendency predicts by the model is rather well respected.

 \section{Conclusion}
  The melting of ice disc on a thermalized  plate allows to uncover a behavior similar to Leidenfrost effet. The change of phase due to the proximity of a thermalised plate generates a film of the transformed phase. The block is then isolated from the plate and ``levitates'' on the lubrication film. We demonstrate that in addition, the ice disc rotates while melting. This phenomenon has been modelled by considering the flow of the water. The water which flows around the ice disc entrains the rotation of the ice disc.

\begin{acknowledgments}
SD acknowledges support from FNRS as Senior Research Associate. This research has been funded by the Interuniversity Attraction Pole Programme (IAP 7/38 MicroMAST) initiated by the Belgian Science Policy Office.
\end{acknowledgments}


\end{document}